\documentclass[aps,prx,twocolumn]{revtex4-1}
\usepackage{graphicx}
\usepackage{amsmath}
\usepackage{amssymb}
\usepackage[english]{babel}
\usepackage{xcolor}
\usepackage{siunitx}

\usepackage[colorlinks=true, pdfstartview=FitV, linkcolor=blue,
citecolor=blue, urlcolor=blue]{hyperref}
\usepackage[capitalise]{cleveref}

\makeatletter
\let\saved@includegraphics\includegraphics
\AtBeginDocument{\let\includegraphics\saved@includegraphics}
\renewenvironment*{figure}{\@float{figure}}{\end@float}
\renewenvironment*{table}{\@float{table}}{\end@float}
\makeatother

\newcommand\T{\rule{0pt}{2.6ex}}       
\newcommand\B{\rule[-1.2ex]{0pt}{0pt}} 

\begin{document}

\title{Strong spin-orbit interaction and $g$-factor renormalization of\\ hole spins in Ge/Si nanowire quantum dots}

\author{F. N. M. Froning$^{1}$}\thanks{These authors contributed equally}
\author{M. J. Ran\v ci\'c$^{1, 2}$}\thanks{These authors contributed equally}
\author{B. Het\'enyi$^{1}$}
\author{S. Bosco$^{1}$}
\author{M. K. Rehmann$^{1}$}
\author{A. Li$^{3}$}
\author{E. P. A. M. Bakkers$^{3}$}
\author{F. A. Zwanenburg$^{4}$}
\author{D. Loss$^{1}$}
\author{D. M. Zumb\" uhl$^{1}$}
\author{F. R. Braakman$^{1}$}\thanks{Author to whom correspondence should be addressed.  Electronic mail: \texttt{floris.braakman@unibas.ch}.}

\affiliation{1: Department of Physics, University of Basel, Klingelbergstrasse 82, 4056 Basel, Switzerland}
\affiliation{2: Total S.A., Nano-INNOV, B\^at .861 8, Boulevard Thomas Gobert, 91120 Palaiseau, France}
\affiliation{3: Department of Applied Physics, Eindhoven University of Technology, P.O. Box 513, 5600 MB Eindhoven, The Netherlands}
\affiliation{4: NanoElectronics Group, MESA+ Institute for Nanotechnology, University of Twente, P.O. Box 217, 7500 AE Enschede, The Netherlands}

\begin{abstract}
The spin-orbit interaction lies at the heart of quantum computation with spin qubits, research on topologically non-trivial states, and various applications in spintronics. 
Hole spins in Ge/Si core/shell nanowires experience a spin-orbit interaction that has been predicted to be both strong and electrically tunable, making them a particularly promising platform for research in these fields.
We experimentally determine the strength of spin-orbit interaction of hole spins confined to a double quantum dot in a Ge/Si nanowire by measuring spin-mixing transitions inside a regime of spin-blockaded transport. We find a remarkably short spin-orbit length of $\sim$\SI{65}{nm}, comparable to the quantum dot length and the interdot distance. 
We additionally observe a large orbital effect of the applied magnetic field on the hole states, resulting in a large magnetic field dependence of the spin-mixing transition energies. Strikingly, together with these orbital effects, the strong spin-orbit interaction causes a significant enhancement of the $g$-factor with magnetic field.
The large spin-orbit interaction strength demonstrated is consistent with the predicted direct Rashba spin-orbit interaction in this material system and is expected to enable ultrafast Rabi oscillations of spin qubits and efficient qubit-qubit interactions, as well as provide a platform suitable for studying Majorana zero modes.
\end{abstract}
\maketitle

\section{Introduction}
The spins of single electrons or holes can be coupled to orbital degrees of freedom through the spin-orbit interaction. In a solid-state environment, this interaction arises from the motion of electrons or holes in electric fields associated with the host lattice atoms, structural or bulk inversion fields, or externally applied electric fields, and its strength can range from a typically small perturbation in the conduction band to a significant effect in the valence band~\cite{WinklerSpinOrbitCoupling2003}. Spin-orbit interaction is particularly useful for fundamental applications in spintronics and quantum information processing with spin qubits~\cite{LossQuantumcomputationquantum1998,HansonSpinsfewelectronquantum2007,Scappuccigermaniumquantuminformation2020}, as it can be employed to realize fast manipulation of spin states purely through electrical means~\cite{GolovachElectricdipoleinducedspinresonance2006,NowackCoherentControlSingle2007}.
For example, Rabi oscillations with frequencies of $\sim$\SI{100}{\mega\hertz} have been obtained for electron spins confined in group III-IV semiconductor nanowires, where the spin-orbit interaction was used to mediate a coupling of the spins to an electrical driving field~\cite{Nadj-PergeSpinorbitqubitsemiconductor2010,vandenBergFastSpinOrbitQubit2013}. Furthermore, spin-orbit interaction provides a promising path towards implementing entangling operations between distant spin qubits, by mediating the coupling of spins to electromagnetic cavity modes~\cite{KloeffelCircuitQEDholespin2013,BurkardSuperconductorsemiconductorhybridcircuit2020} or floating gate architectures~\cite{TrifunovicLongDistanceSpinSpinCoupling2012}. An important advantage of using spin-orbit interaction for these purposes is that it requires no additional on-chip components such as micromagnets.

The emergence of Majorana zero modes in semiconductor nanowires relies on the presence of a strong spin-orbit interaction~\cite{LutchynMajoranaFermionsTopological2010,OregHelicalLiquidsMajorana2010, AliceaMajoranafermionstunable2010,KlinovajaCompositeMajoranafermion2012}. When combined with conventional bulk s-wave superconductivity, induced in the nanowire through proximitization, and with a Zeeman field, sufficiently strong spin-orbit interaction results in an effective 1D p-wave superconductor supporting Majorana zero modes. Such Majorana zero modes are of fundamental interest since they exhibit exotic non-Abelian statistics and hold great promise to realize quantum computation with topological protection from decoherence~\cite{KitaevFaulttolerantquantumcomputation2003}. The strength of the spin-orbit interaction sets the range of Zeeman energies in which a topologically non-trivial phase exists together with a sufficiently large superconducting gap, making a strong spin-orbit interaction essential for experimental studies~\cite{MaierMajoranafermionsGe2014}.

Hole spins in semiconductor nanostructures can experience a spin-orbit interaction many times stronger than for electron spins~\cite{WinklerSpinOrbitCoupling2003, BulaevSpinRelaxationDecoherence2005, BulaevElectricDipoleSpin2007}. In particular, a strong and electrically tunable direct Rashba spin-orbit interaction arises for holes confined in one-dimensional Ge- or Si-based nanostructures~\cite{KloeffelStrongspinorbitinteraction2011,KloeffelDirectRashbaspinorbit2018}. The direct Rashba spin-orbit interaction results from direct dipolar coupling of holes to an external electric field, in combination with mixing of heavy and light hole states due to confinement to one dimension. This interaction is estimated to be 10-100 times stronger than the conventional Rashba-type spin-orbit interaction for electrons or holes.

\begin{figure}[t]
	\includegraphics[width=0.5\textwidth]{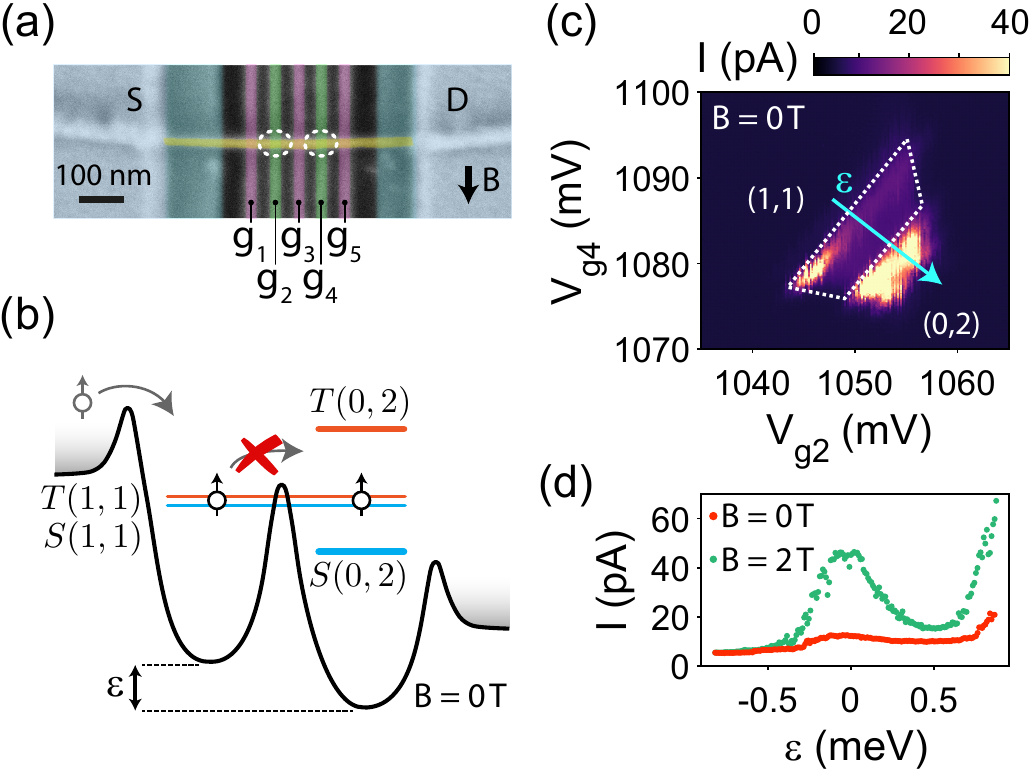}
	\caption{Device and Pauli spin blockade. (a) False-colour scanning elecron micrograph of the device, used for all the measurements of this work. The finger gates $g_\text{1-5}$ (red: barrier gates, green: plunger gates) are biased with positive voltages $V_\text{g1-5}$ in order to create a double quantum dot in the Ge/Si core/shell nanowire (yellow). The source (S) and drain (D) contacts are defined on either side of the nanowire. Dashed ellipses indicate the approximate locations of the two quantum dots. (b) Schematic illustration of Pauli spin blockade, with zero magnetic field. When the double dot is occupied by holes in a triplet $(1,1)$ state, the current is blocked until mixing with a singlet state takes place. The double dot detuning is indicated by $\varepsilon$. (c) Bias triangles taken at $V_\text{SD}$\,=\,\SI{2}{mV} showing signatures of Pauli spin blockade, through a suppression of current, in the area delineated by the dashed white lines. The blue arrow indicates the direction of the detuning axis. (d) Current as a function of detuning, swept along the arrow in (c), without (red) and with (green) applied magnetic field.
}
	\label{fig1}
\end{figure}
Such a strong spin-orbit interaction would enable pushing spin qubit Rabi frequencies into the GHz regime~\cite{KloeffelCircuitQEDholespin2013}, an order of magnitude higher than recently demonstrated with hole spin qubits~\cite{Watzingergermaniumholespin2018, HendrickxFasttwoqubitlogic2020, Hendrickxsingleholespinqubit2019}, and state-of-the-art electron-based spin qubits~\cite{vandenBergFastSpinOrbitQubit2013,YonedaFastElectricalControl2014, Yonedaquantumdotspinqubit2018}. 
Moreover, a large electrical tunability of spin-orbit interaction strength promises exquisite control over qubit coherence and manipulation speeds, providing a gate-controlled $\textit{ON}/\textit{OFF}$ switch of the coupling to electrical environmental degrees of freedom, which could be used to, on the one hand, maximize the coupling to microwave drive fields and, on the other hand, minimize the coupling to charge noise. Such controllable coupling would make it possible to combine ultrafast qubit operations with long coherence times. Furthermore, such electrical tunability can be used to control the localization length of Majorana zero-modes confined to each end of a nanowire~\cite{MaierMajoranafermionsGe2014}, creating the possibility of electrically performing topologically non-protected operations on Majorana zero-modes.

Due to the tunable nature of the spin-orbit interaction, the magnitude of the $g$-factor of hole spins in Ge/Si nanowires can be modulated over a large range using applied electric fields~\cite{MaierTunablefactorphononmediated2013, BraunsElectricfielddependentfactor2016}. This feature enables local control over the Zeeman energy and allows to tune the energy of a qubit relative to a spin resonance driving field, or to a microwave cavity mode, making it possible to selectively address individual qubits in a multi-qubit device. Furthermore, in addition to strong and tunable spin-orbit interaction, hole spins in Ge/Si nanowires combine several other features that make them amenable for implementation of high-quality qubits. Hyperfine-induced decoherence is expected to be strongly suppressed, since holes have a p-type Bloch function, which has zero overlap with lattice nuclear spins~\cite{FischerSpindecoherenceheavy2008}. Furthermore, both Ge and Si have a low natural abundance of isotopes with non-zero nuclear spins ($^{29}\text{Si} <5\%$, $^{29}\text{Ge}<8\%$), which can be made vanishingly small through isotopic purification. Finally, in contrast to electrons, holes in Ge and Si do not experience valley degeneracy,  which for electron spins in Si-based devices can have a detrimental effect on qubit relaxation times~\cite{YangSpinvalleylifetimessilicon2013}.\\

Here, we investigate the spin-orbit interaction of hole spins confined in a double quantum dot defined electrostatically in a Ge/Si core/shell nanowire~\cite{FroningSingledoubletriple2018a, BraunsHighlytuneablehole2016}. We use mixing of singlet and triplet spin states detected through lifting of Pauli spin blockade~\cite{OnoCurrentRectificationPauli2002a, KoppensControlDetectionSingletTriplet2005,Nadj-PergeDisentanglingeffectsspinorbit2010, BraunsAnisotropicPaulispin2016, ZarassiMagneticfieldevolution2017} to perform spectroscopy on the effectively doubly occupied double dot. Notably, we also find a large orbital effect of the magnetic field. We have developed a spectroscopic model, which fully takes into account these orbital effects, allowing to independently determine the Land\'e $g$-factor, the interdot tunnel coupling strength, and the strength of the spin-orbit interaction in this device. We find a particularly strong spin-orbit interaction, with a spin-orbit length of the same order as the dot size. Such a regime of strong spin-orbit interaction is expected to exhibit effects~\cite{TrifSpindynamicsInAs2008, DmytrukRenormalizationquantumdot2018} typically not observed in experiments with quantum dots. Specifically, it causes a renormalization of the $g$-factor, which we find here to lead to a Zeeman energy that is a non-linear function of the applied magnetic field.

\begin{figure}[t]
	\includegraphics[width=0.48\textwidth]{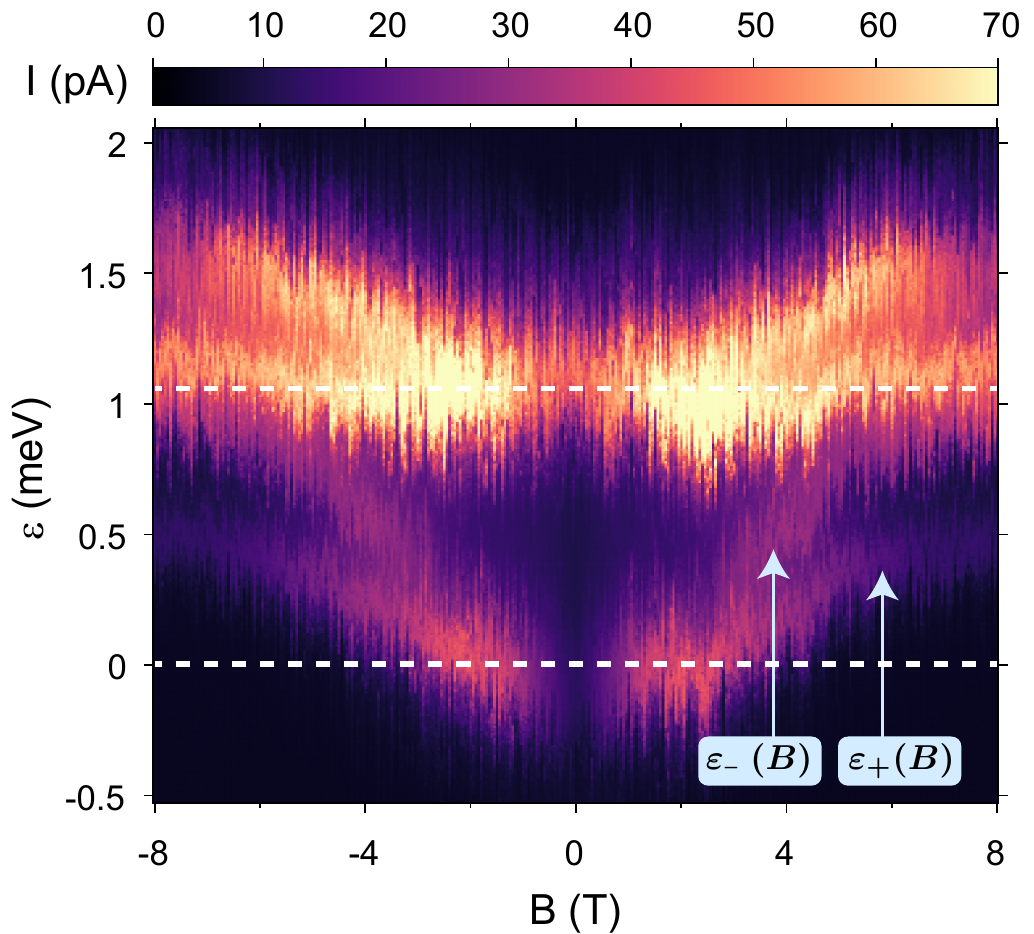}
	\caption{Measured leakage current as a function of magnetic field for detunings covering the entire bias triangle, as shown by the arrow in Fig.~\ref{fig1}(c). The dashed white lines delineate the spin-blockaded region also shown in Fig.~\ref{fig1}(c). Here, $V_\text{g3}$\,=\,\SI{3820}{\milli\volt}.}
	\label{fig2}
\end{figure}
\section{Device and measurement setup}
The device we use consists of a single Ge/Si core/shell nanowire deterministically placed on top of five finger gates, which are equally spaced with a pitch of $\SI{50}{\nano\meter}$ (see Fig.~\ref{fig1}(a)). The nanowire has an overall radius of $\SI{11}{\nano\meter}\pm \SI{2}{\nano\meter}$, as determined through atomic force microscopy, and a nominal Si shell thickness of $\SI{2.5}{\nano\meter}$. A $\SI{20}{\nano\meter}$ thick layer of $\text{Al}_2\text{O}_3$ in between gates and nanowire serves as electrical insulation. Electrical contact to the nanowire is made through two Ti/Pd contact pads, defined on either side of the nanowire. For more details of the device, see Froning et al.~\cite{FroningSingledoubletriple2018a}. Previously, we have shown a large degree of control over the formation of quantum dots in such devices, which can be tuned over hundreds of charge transitions down to the few-holes occupation regime~\cite{BraunsHighlytuneablehole2016,FroningSingledoubletriple2018a}. Here, we form a tunnel-coupled double quantum dot by applying positive voltages to the finger gates $g_\text{1}-g_\text{5}$ that locally deplete the nanowire hole gas~\cite{LuOnedimensionalholegas2005}. We use the contact pads to apply a source-drain voltage bias of $V_\text{SD}$\,=\,\SI{2}{\milli\volt} across the nanowire and to measure the current flowing through the double dot. An external magnetic field is applied in the sample plane, perpendicular to the major axis of the nanowire, as indicated in Fig.~\ref{fig1}(a). All measurements were taken at a temperature of \SI{1.4}{K}.

\section{Double quantum dot and Pauli spin blockade}
We tune the double dot to an effective occupation of two holes and study the transport cycle $(0,1)\rightarrow(1,1)\rightarrow(0,2)\rightarrow(0,1)$ in a Pauli spin blockade~\cite{OnoCurrentRectificationPauli2002a,HansonSpinsfewelectronquantum2007} configuration (see Fig.~\ref{fig1}(b)). Here the first and second numbers refer to the effective hole occupation of the left and right dot, respectively. Transport in this regime is subject to a spin selection rule imposed by the Pauli exclusion principle: interdot transitions $(1,1)\rightarrow(0,2)$ are blocked for spin triplet states ($|T_{\downarrow\downarrow}\rangle,|T_0\rangle, |T_{\uparrow\uparrow}\rangle$, with spin quantum numbers $s$\,=\,1 and $m_s$\,=\,-1,\,0,\,+1), since the $|T(0,2)\rangle$ states are energetically inaccessible. In contrast, interdot transitions are energetically allowed for holes in a spin singlet state ($|S\rangle$, $s=m_s=0$). Therefore, when a triplet $(1,1)$ state gets occupied, current through the double dot is blocked, until mixing with a singlet state takes place.

We exploit such spin-selective transport as a read-out method allowing us to distinguish spin states~\cite{OnoCurrentRectificationPauli2002a}. Fig.~\ref{fig1}(c) shows a measurement of the current through the double dot as a function of the voltage on gates $g_2$ and $g_4$, taken at zero magnetic field. We identify the area of reduced current, enclosed by the dashed line in Fig.~\ref{fig1}(c), as a signature of spin blockade. Consistently, for opposite $V_\text{SD}$, we obtain a larger current (not shown). Furthermore, as can be seen in the traces of Fig.~\ref{fig1}(d), the blockade is lifted at a finite magnetic field, resulting in an increased current. Even when in a triplet state, transport can become unblocked~\cite{OnoCurrentRectificationPauli2002a} through various spin-mixing mechanisms that coherently or incoherently couple triplet and singlet states. Possible spin-mixing mechanisms are based on hyperfine interactions with the nuclear spin bath of the host lattice~\cite{KoppensControlDetectionSingletTriplet2005, DanonPaulispinblockade2009, Nadj-PergeDisentanglingeffectsspinorbit2010}, spin-flip cotunneling~\cite{QassemiStationaryTransientLeakage2009,CoishLeakagecurrentlineshapes2011,BiesingerIntrinsicMetastabilitiesCharge2015,BraunsAnisotropicPaulispin2016}, $g$-factor differences in the double quantum dot, and spin-orbit interaction~\cite{DanonPaulispinblockade2009, Nadj-PergeDisentanglingeffectsspinorbit2010, LiPauliSpinBlockade2015, BraunsAnisotropicPaulispin2016,ZarassiMagneticfieldevolution2017}. The dominant spin-mixing mechanism can be investigated by leakage currents in Pauli spin blockade.

\section{Lifting of Pauli Spin Blockade}
We study the lifting of spin blockade in more detail, focussing on the dependence of the resulting leakage current on double-dot detuning $\varepsilon$, magnetic field $B$, and interdot tunnel coupling strength $t_c$. Fig.~\ref{fig2} shows a measurement of the current through the double dot as a function of magnetic field $B$ and detuning $\varepsilon$. The latter is swept over the entire bias triangle, by changing $V_\text{g2}$ and $V_\text{g4}$ following the arrow in Fig.~\ref{fig1}(c). 
The white dashed lines in Fig.~\ref{fig2} indicate the spin-blockaded regime $0<\varepsilon<\varepsilon_\Delta$, with $\varepsilon_\Delta\approx$\,\SI{1}{meV} the detuning for which states with one hole in the first orbital excited state becomes energetically available. 
For detunings exceeding $\varepsilon_\Delta$, we observe features with a significantly increased current. We attribute these features to spin-flip transitions involving a higher orbital state, i.e. either $|T_{\uparrow\uparrow,\downarrow\downarrow}(1,1)\rangle - |S_\Delta(0,2)\rangle$, or $|S(1,1)\rangle-|T_{\uparrow\uparrow,\downarrow\downarrow}(0,2)\rangle$ transitions, where $|S_\Delta\rangle$ refers to a singlet state with one hole in the orbital ground state and one hole in the first orbital excited state. Note that also spin-conserving $|T(1,1)\rangle-|T(0,2)\rangle$ transitions can take place for these detunings, but these transitions would not exhibit multiple peaks at finite magnetic field, since they do not exhibit a Zeeman splitting. Remarkably, we find that in our experiment transitions that do not conserve spin have a higher amplitude than transitions that do conserve spin, as discussed later.\\

Here we are interested in the spin-blockaded region and in the remaining part we focus on the features between the white lines in Fig.~\ref{fig2}.
In this range of detuning, we see a markedly increased current that correspond to lifting of Pauli spin blockade. These leakage current features form the main topic of this work. We can make two important observations: \textit{1)} for a given sign of $B$, the leakage current is maximum along two curves as a function of $\varepsilon$ and $B$, marked $\varepsilon_{\pm}(B)$ in Fig.~\ref{fig2}; \textit{2)} around zero magnetic field the leakage current is suppressed. These observations form the starting point in identifying the triplet-singlet transitions underlying the leakage current along $\varepsilon_{\pm}(B)$, as well as the spin-mixing mechanism. 

As explained in more detail in Section~\ref{sec:model}, the position of the two curves as a function of detuning and magnetic field allows us to assign them to $|T_{\uparrow\uparrow,\downarrow\downarrow}(1,1)\rangle-|S(0,2)\rangle$ transitions. These transitions occur at different detuning depending on the magnetic field, due to an increase in Zeeman splitting, as well as orbital effects of the magnetic field. As shown in the next section, we identify spin-orbit interaction as the dominant spin-mixing mechanism by evaluating the magnetic field-dependent intensity of these transitions.

\begin{figure*}[t]
	\includegraphics[width=1\textwidth]{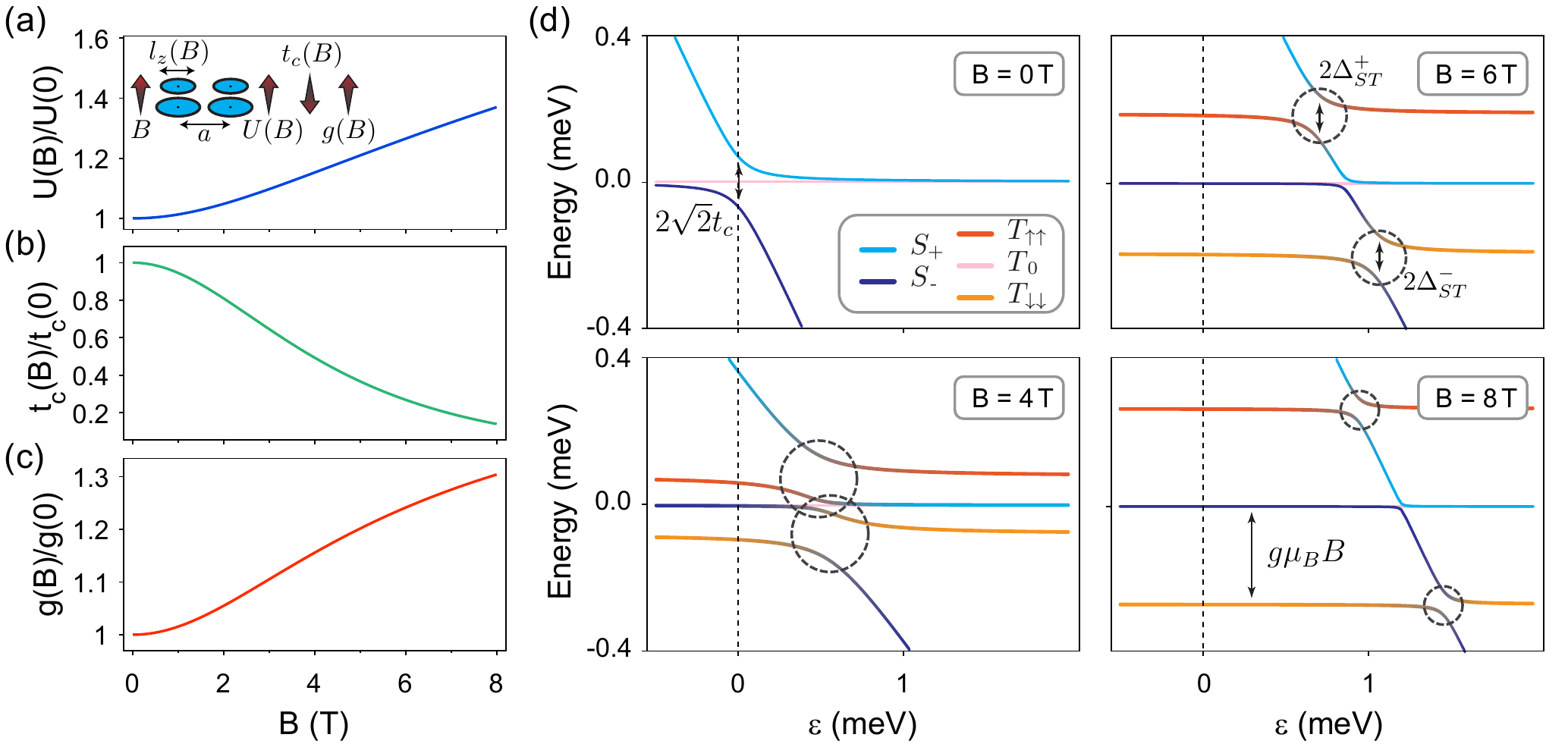}
	\caption{Magnetic-field dependencies. (a) Calculated magnetic field dependence of the addition energy $U$ (See Eq.~\eqref{eq:U-charging} of Appendix~\ref{app:B}). Inset: Schematic illustration of the effect of increasing magnetic field $B$ on dot size and separation leading to the observed changes in $U$, $t_c$ and $g$. Quantities change qualitatively with $B$ as indicated by the arrows. (b) Calculated magnetic field dependence of the spin-conserving tunnel coupling $t_c$ (see Eq.~\eqref{eq:tunneling} of Appendix~\ref{app:B}). (c) Calculated magnetic field dependence of the $g$-factor (see Eq.~\eqref{eq:g-factor}). For the plots in (a)-(c), the relevant parameters correspond to those of the measurement of Fig.~\ref{fig2}. (d) Double dot energy level diagrams for different values of the magnetic field. For $B$\,=\,\SI{0}{T}, the spin-conserving tunnel coupling $t_c$ is maximum and there is no singlet-triplet mixing due to spin-orbit interaction. For large enough magnetic fields ($B>\tilde{B}$), avoided crossings (highlighted by dashed circles) appear when the triplet $(1,1)$ states cross a singlet state with $(0,2)$ component, corresponding to spin-flip tunneling due to spin-orbit interaction. The size of all avoided crossings becomes smaller with increasing magnetic field, as can be understood from (b) and Eq.~\eqref{eq:DeltaST}. Moreover, due to the magnetic field dependence of the addition energy $U$ (see (a)), as well as the Zeeman energy, all avoided crossings move to higher detuning with magnetic field. Parameters used to plot the diagrams were extracted from the data set shown in Fig.~\ref{fig2}, using the model described in the text.}
	\label{fig3}
\end{figure*}

\section{Possible Spin-mixing mechanisms}
We now discuss the origin of the spin mixing leading to the observed lifting of spin blockade by considering the dependence of possible spin-mixing mechanisms on the magnetic field and detuning. In particular, the zero-field gap can be attributed to spin-orbit interaction, which is not effective at $B$\,=\,\SI{0}{T} due to time-reversal invariance~\cite{GolovachSpinrelaxationsinglettriplet2008, DanonPaulispinblockade2009}, but becomes important at finite $B$~\cite{SchreiberCouplingartificialmolecular2011, BraakmanDynamicsspinflipphotonassisted2014}. 
Furthermore, for $\varepsilon$\,=\,0 and $|B|$ smaller than a characteristic field $\tilde{B}$, the triplet $(1,1)$ states lie within the $|S(1,1)\rangle-|S(0,2)\rangle$ avoided crossing, at which point spin-orbit interaction does not couple them efficiently to the singlet states.

Spin-flip cotunneling can also lead to dips or peaks in the leakage current around $B$\,=\,\SI{0}{T}. Such spin-flip cotunneling involves the exchange of a hole spin with one of the lead reservoirs through a process involving a virtual intermediate state, which can lead to decay of the triplet $(1,1)$ to a singlet state. Such cotunneling can result in a leakage current peak at $B$\,=\,\SI{0}{T} that exists for $\varepsilon$\,=\,0, as well as for values of $\varepsilon$ up to $\varepsilon_\Delta$. A shallow zero-field dip can also result from cotunneling, when the temperature $T$ is small compared to $t_c$ ~\cite{CoishLeakagecurrentlineshapes2011, QassemiStationaryTransientLeakage2009}.
However, the data presented in Fig.~\ref{fig2} shows a deep zero-field gap and our operating temperature of $\SI{1.4}{\kelvin}$ is, as will be shown later, comparable to $t_c$.
We therefore rule out spin-flip cotunneling as the dominant spin-mixing mechanism in our measurements. 

Furthermore, fluctuating polarizations of the nuclear spin bath in the double dot can result in triplet-singlet mixing~\cite{KoppensControlDetectionSingletTriplet2005, JouravlevElectronTransportDouble2006, Nadj-PergeDisentanglingeffectsspinorbit2010}. However, as mentioned in the introduction, hyperfine interaction is expected to be very small for hole spins in Ge- and Si-based devices. Moreover, this mechanism is only effective for values of $B$ up to the root mean square value of nuclear field fluctuations, which we estimate to be $<\!\SI{1}{\milli\tesla}$ in our system~\cite{KoppensControlDetectionSingletTriplet2005}. Most notably, in contrast to what we observe, this spin-mixing mechanism should result in a leakage current peak~\cite{KoppensControlDetectionSingletTriplet2005, Nadj-PergeDisentanglingeffectsspinorbit2010} around $B$\,=\,\SI{0}{T} for $\varepsilon$ up to $\varepsilon_\Delta$. 

Finally, differences in $g$-factor between the two dots need to be considered. The effective $g$-factor for holes in Ge/Si nanowires can depend sensitively on the electric field~\cite{MaierTunablefactorphononmediated2013}, confinement potential~\cite{TrifSpindynamicsInAs2008, DmytrukRenormalizationquantumdot2018}, and hole occupation number. At finite field, such a $g$-factor difference will mix the $|T_0(1,1)\rangle$ and $|S(1,1)\rangle$ states, thus leading to an additional resonance of the leakage current. However, such $|T_0(1,1)\rangle - |S(1,1)\rangle$ mixing would not result in the two separated curves of increased current that we observe, but instead provide a background leakage current in the detuning range considered, with no magnetic field dependence. Note further that such mixing is suppressed as $|T_0(1,1)\rangle$ is split off from the singlet by the exchange energy. 

In conclusion, we identify spin-orbit interaction as the dominant spin-mixing mechanism responsibe for the observed leakage current. In a double quantum dot, spin-orbit interaction can flip the spin of a hole tunneling between the quantum dots. This enables triplet-singlet mixing, when these states are aligned in energy, which can effectively lift Pauli spin blockade. As shown in the next section, we can explain the spectroscopy of the observed leakage current using this mechanism.

\section{Model of the two transitions}
\label{sec:model}
Here, we present an analytical model that takes into account non-spin-conserving interdot tunneling and its dependence on magnetic field and detuning. Our model agrees very well with the data and accurately reproduces the field-dependence of the two observed transitions shown in Fig.~\ref{fig2}, allowing us to identify them as $|T_{\uparrow\uparrow,\downarrow\downarrow}\rangle - |S\rangle$ transitions.

As mentioned before, we assume that the spin-blockade and its lifting can be understood in terms of an effectively doubly-occupied double dot.
When the spin-conserving interdot tunnel coupling $t_c$ is finite, the singlet states $|S(0,2)\rangle$ and $|S(1,1)\rangle$ are coupled, giving rise to two new eigenstates we refer to as the lower and higher hybridized singlet states, $|S_-\rangle$ and $|S_+\rangle$, respectively~\cite{StepanenkoSinglettripletsplittingdouble2012}. These hybridized singlets are defined as $|S_-\rangle = \sin(\theta/2)|S(1,1)\rangle - \cos(\theta/2)|S(0,2)\rangle$ and $|S_+\rangle = \cos(\theta/2)|S(1,1)\rangle + \sin(\theta/2)|S(0,2)\rangle$, with the mixing angle $\theta$ being a function of detuning $\varepsilon$ and $t_c$ (see Eq.~\eqref{Eq:theta} for the full expression of $\theta$). 
The $|S_\pm\rangle$ states exhibit an avoided crossing around $\varepsilon$\,=\,0 with a gap of $2\sqrt{2}t_c$, as shown in Fig.~\ref{fig3}(d).
Importantly, the proportion of $|S(0,2)\rangle$ and $|S(1,1)\rangle$ present in each of the $|S_\pm\rangle$ states depends on the detuning.

In the presence of spin-orbit interaction, spin-flip tunneling couples the
$|T_{\uparrow\uparrow,\downarrow\downarrow}(1,1)\rangle$ states with the two hybridized $|S_\pm\rangle$ states, due to the $|S(0,2)\rangle$ content of the latter. The coupling strength of this spin-flip tunneling is given by the strength of the spin-conserving tunnel coupling as well as the strength of the spin-orbit interaction and can be written as $t_{so}= t_c \tan\big(a/\lambda_{so}\big)$ (see Appendix~\ref{app:B} for derivation), with $a$ the interdot distance and $\lambda_{so}$ the spin-orbit length (defined by $\pi\lambda_{so}/2$ being the distance a hole has to travel for spin-orbit interaction to induce a $\pi$-rotation of its spin state). 

This coupling leads to avoided crossings when the energies of the $|T_{\uparrow\uparrow,\downarrow\downarrow}\rangle$ states exactly match the energies of the $|S_\pm\rangle$ states, as illustrated in the energy level diagrams in Fig.~\ref{fig3}(d). The leakage current is maximum for those values of the detuning where the triplet-singlet avoided crossings occur, which can be written as: 
\begin{equation}
\label{eq:epsa}
\varepsilon_{\pm}(B) = U(B)-U(0) \pm \left(\frac{2 t_c^2(B)}{g(B)\mu_{\rm B} B} -g(B)\mu_{\rm B} B\right) \ .
\end{equation}
Here the indices $+$ and $-$ correspond to the $|T_{\uparrow\uparrow}\rangle -|S_+\rangle$ and $|T_{\downarrow\downarrow}\rangle - |S_-\rangle$ transitions, respectively. Furthermore, $\mu_{\rm B}$ is the Bohr magneton, $g$ the $g$-factor in the dot, and $U$ the single dot addition energy. 
Eq.~\eqref{eq:epsa} describes the evolution of spin-blockade leakage current with magnetic field shown in Fig.~\ref{fig2} between the white dashed lines, with $\varepsilon_{\pm}(B)$ giving the detunings of the resonant peaks of the two features as a function of magnetic field.\\

In order to explain the precise magnetic field dependence of $\varepsilon_{\pm}(B)$, we need to take into account effects that rely on the magnetic field changing the size of the hole orbitals. In the experiment, the magnetic field is oriented perpendicular to the principal nanowire axis and is varied over a wide range of amplitudes ($\SI{-8}{\tesla}\le B \le \SI{8}{\tesla}$), making such orbital effects significant in this system. 

Remarkably, this turns the spin-conserving tunnel coupling $t_c$, the addition energy $U$ and the $g$-factor into quantities that all depend on the magnetic field (see inset Fig.~\ref{fig3}(a)). Such effects are usually dealt with only qualitatively, even though their relative magnitude can be quite large. Here, we take these effects fully into account in our spectroscopic model, enabling us to quantify the $g$-factor and the spin-orbit length in our device.\\

To derive the functional dependence of these quantities on $B$, we start from the Hund-Mulliken theory of atomic orbitals and we assume harmonic confinement in all three directions. By considering an anisotropic 3-dimensional oscillator, we model the effects of a confinement potential that is smoother (sharper) in the direction along (perpendicular to) the nanowire as well as the strain-induced anisotropy of the effective mass ~\cite{KloeffelDirectRashbaspinorbit2018}.
The hole wavefunctions in each dot are squeezed by the magnetic field and as a result the spin-conserving tunneling $t_c(B)$ is reduced at large fields while the single-dot addition energy $U(B)$ is enhanced, as shown schematically in the inset of Fig.~\ref{fig3}(a). The explicit dependencies of $t_c(B)$ and $U(B)$ on magnetic field are given in Eqs.~\eqref{eq:tunneling} and \eqref{eq:U-charging} of Appendix~\ref{app:B}, and are plotted in Fig.~\ref{fig3}(a), (b).\\

The detunings at which the $|T_{\uparrow\uparrow,\downarrow\downarrow}\rangle - |S_\pm\rangle$ avoided crossings appear also depend on the Zeeman splitting $E_\text{Z}$ of the $|T_{\uparrow\uparrow,\downarrow\downarrow}\rangle$ states with respect to the singlets. Usually, the Zeeman splitting is a linear function of the magnetic field, which can be written in terms of the $g$-factor as $E_Z = g\mu_B|B|$. However, strong spin-orbit interaction can renormalize the $g$-factor~\cite{TrifSpindynamicsInAs2008, DmytrukRenormalizationquantumdot2018} when the size of the quantum dot is changed. 
In our case, the magnetic field changes the dot size through orbital effects, leading to a dependence of the $g$-factor on the magnetic field and turning the Zeeman energy into a non-linear function of the magnetic field.

The shrinking of the dot with increasing magnetic field causes the $g$-factor to be enhanced at large values of the magnetic field and we can write~\cite{TrifSpindynamicsInAs2008, DmytrukRenormalizationquantumdot2018} 
\begin{equation}
\label{eq:g-factor}
g(B) = g_0e^{-\frac{l_\parallel^2}{\lambda_{so}^2}\left(1+ \frac{B^2}{B_0^2}\right)^{-1/2}} \ ,
\end{equation} 
where $g_0$ is the $g$-factor without the spin-orbit-induced renormalization. Furthermore, $l_{\parallel}$ is the field-independent harmonic length of the hole wavefunction ($l_{\parallel} = l_z(B=0)$, with $l_z$ being the dot confinement length along the wire) and $B_0$ is a characteristic magnetic field that depends on the average confinement strength in the directions perpendicular to the field. See Appendix~\ref{app:A} for the precise definition of these quantities. Fig.~\ref{fig3}(c) shows a plot of Eq.~\eqref{eq:g-factor}, using the values of $l_{\parallel}$ and $B_0$ extracted from the measurement of Fig.~\ref{fig2}. We stress that the magnetic-field dependence of the $g$-factor in Eq.~\eqref{eq:g-factor} is a direct consequence of the strong spin-orbit interaction in the nanowire and it vanishes when the spin-orbit length $\lambda_{so}$ is much larger than the dot size, which is typically the case for quantum dot systems that have been experimentally realized thus far.

As will be shown in the next section, when taking into account the magnetic field dependence of $U$, $t_c$, and $g$, the resonant positions $\varepsilon_\pm(B)$ of the $|T_{\uparrow\uparrow,\downarrow\downarrow}\rangle - |S_\pm\rangle$ transitions given by Eq.~\eqref{eq:epsa} closely reproduce the evolution of the two features of spin blockade leakage current of Fig.~\ref{fig2} as a function of magnetic field and detuning.

\begin{figure*}[t]
	\includegraphics[width=1\textwidth]{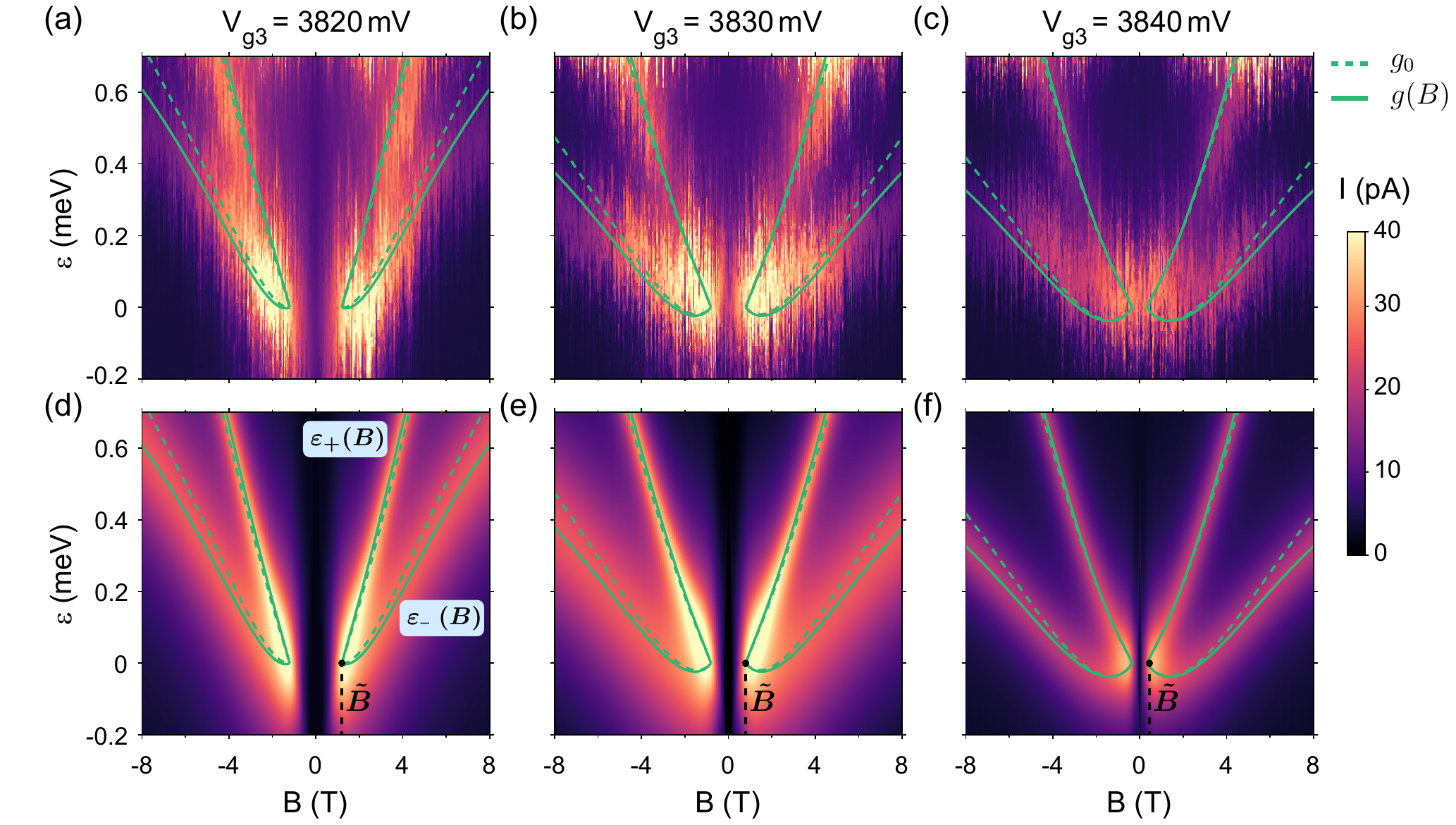}
	\caption{Spectroscopy measurements and modelling. (a)-(c) Measured leakage current as a function of magnetic field and detuning $\varepsilon < \varepsilon_\Delta$, for $V_\text{g3} =$ 3820, 3830, and \SI{3840}{\milli\volt}. The green curves are fits of each data set to Eq.~\eqref{eq:epsa}, with (solid) and without (dashed) taking into account $g$-factor renormalization with magnetic field. (d)-(f), Simulated leakage current as a function of magnetic field and detuning. Here, we used the model discussed in Sections~\ref{sec:model}-\ref{sec:spinorbitlength} of the main text, with relevant parameters determined from fits of the data shown in (a)-(c). The green curves are identical to the curves in (a)-(c).
}
	\label{fig4}
\end{figure*}

\section{Varying the strength of interdot tunnel coupling}
\label{sec:differenttunnelcouplings}
To demonstrate the versatility of our model we now explore the influence of varying the voltage $V_\text{g3}$ on the middle gate on the leakage current. The main expected effects are a change in the interdot tunnel coupling $t_c$ and a change in the dot confinement. Figs.~\ref{fig4}(a)-(c) show measurements similar to that of Fig.~\ref{fig2}, for three values of $V_\text{g3}$ (see Fig.~S1 of the Supplemental Material for extended data sets). Comparing the three data sets, we see that an increase of $V_\text{g3}$ leads to a closing of the zero-field gap.
As discussed before, Pauli spin blockade only becomes lifted through spin-orbit interaction for magnetic fields above a critical value. This critical field $\tilde{B}$ can be written as
\begin{equation}
\label{eq:tildeB}
\tilde{B}=\frac{\sqrt{2}}{\mu_{\rm B}}\frac{t_c(\tilde{B})}{g(\tilde{B})} \ ,
\end{equation} 
where we include the magnetic field dependence of $t_c$ and $g$. When $|B| = \tilde{B}$, the Zeeman energy matches the size of half of the avoided crossing given by $t_c$. At this point, $\varepsilon_-(B)=\varepsilon_+(B)$ (see Eq.~\eqref{eq:epsa}) and both $|T_{\uparrow\uparrow,\downarrow\downarrow}(1,1)\rangle - |S_\pm\rangle$ transitions become possible at $\varepsilon\approx 0$ (see Fig.~\ref{fig3}(d) and Fig.~\ref{fig4}(a)). For $|B| < \tilde{B}$, each of the singlet-triplet avoided crossings occurs at detunings where the involved $|S_\pm\rangle$ states are mostly composed of $|S(1,1)\rangle$, which does not couple to $|T_{\uparrow\uparrow,\downarrow\downarrow}(1,1)\rangle$ through spin-orbit interaction, leading to a gap in leakage current with characteristic width $\tilde{B}$ around zero magnetic field.\\

By increasing $V_\text{g3}$, we reduce $t_c$ and from Eq.~\eqref{eq:tildeB} it follows that spin blockade can be lifted at smaller magnetic fields. This moves the points of emergence of $\varepsilon_\pm(B)$ for both magnetic field polarities closer together and effectively reduces the width of the zero-field gap of leakage current, in accordance with the observations. In Fig.~\ref{fig4}(a)-(c), we can clearly see this reduction of the zero-field gap (indicated with $\tilde{B}$) when the middle gate voltage $V_\text{g3}$ is increased. Using Eq.~\eqref{eq:tildeB}, we extract the ratio $t_c/g$ at the critical field $\tilde{B}$ for each data set. When the magnetic field is not much larger than $\tilde{B}$, we neglect as a first approximation the variation of $t_c(B)$ and $g(B)$ from their value at $\tilde{B}$, see Figs.~\ref{fig3}(a) and (c), and so using Eq.~\eqref{eq:epsa} we deduce $t_c(\tilde{B})$ and $g(\tilde{B})$ from the relative position of the resonant peaks. Values of $\tilde{B}$, $t_c(\tilde{B})$, and $g(\tilde{B})$ extracted in this way for the three data sets of Fig.~\ref{fig4} are listed in Table~\ref{tab}.\\

By taking into account the orbital effects, our model allows us to explain the main features of the resonances at low magnetic fields.
By linearly expanding the single-dot addition energy in the vicinity of the critical field, $U(B)\approx U(\tilde{B}) + U'(\tilde{B})\big(B - \tilde{B})$, we can approximate $\varepsilon_-(B)\approx \varepsilon(\tilde{B})+ \big(U'(\tilde{B})+g(\tilde{B})\mu_{\rm B}\big)\big(B - \tilde{B})$, reproducing the approximately linear dependence of the upper resonance on magnetic field seen in Fig.~\ref{fig4}. On the other hand, in the expression of the $\varepsilon_+(B)$ resonant peak the term linear in $B$ is smaller and the $1/B$ term gives a significant contribution, leading to a less pronounced shift in detuning, especially at low magnetic field. Although the $1/B$ term is proportional to the tunnel coupling, its effect is counter-intuitively more pronounced in Fig.~\ref{fig4}(c), because here Pauli spin blockade is lifted at lower magnetic fields.

To characterize the overall magnetic field dependence of the leakage current, we now find $\varepsilon_{\pm}(B)$ for each data set by fitting to Eq.~\eqref{eq:epsa}. The green curves in Fig.~\ref{fig4} are plots of $\varepsilon_\pm(B)$ with (solid) and without (dashed) taking into account the renormalization of the $g$-factor given by Eq.~\eqref{eq:g-factor}.
The additional features at larger magnetic fields, such as the bending of the $\varepsilon_+(B)$ curve, are captured by the model by considering the function $U(B)$ beyond the linear approximation, as well as the renormalization of the $g$-factor due to spin-orbit interaction.
We see that the enhancement of the $g$-factor captured by Eq.~\eqref{eq:g-factor} is quite important for large magnetic fields, where it causes a sizeable bending of the resonant peaks (see also Fig.~S2 of the Supplemental Material). Including the renormalized $g$-factor gives much better agreement with the measurements over the whole range of magnetic field values. 

In order to calculate the renormalized $g$-factor using Eq.~\eqref{eq:g-factor}, we estimate the dot confinement length  $l_\parallel=\sqrt{\hbar/(m_\parallel\omega_\parallel)}$, which depends on the confinement energy $\omega_\parallel$ and on the effective mass $m_\parallel$ along the nanowire. We determine $\hbar\omega_\parallel\sim 1\ \text{ meV}$ from measurements of the double dot charge stability diagram and assume $m_\parallel\sim$\,0.05\,$m_0$ (here $m_0$ is the bare electron mass). This choice of $m_\parallel$ is justified by the fact that we still measure a non-zero current even at $|B|=\SI{8}{\tesla}$. If the effective mass along the nanowire growth direction would be smaller, the orbital effects would shrink the wavefunction to the extent that the interdot tunnel coupling would vanish at \SI{8}{\tesla}. For our experiment, we determine $l_\parallel\approx {39-45}$ nm for the range of $V_\text{g3}$ used here. All the parameters extracted from our analysis for the three datasets are reported in Table~\ref{tab}. These values capture the qualitative trend expected: when the voltage $V_\text{g3}$ is increased, the hole wavefunctions become more separated and squeezed, causing a reduction of the tunneling energy $t_c$ and an enhancement of the $g$-factor because of the strong spin-orbit interaction, as described by Eq.~\eqref{eq:g-factor}.
As shown in the next section, our model allows us to extract the spin-orbit length for each measurement. The model color plots shown in Fig.~\ref{fig4}(d)-(f) take into account the extracted values of the spin-orbit length, allowing a full reconstruction of the leakage current in very good agreement with the measurements.

\section{Spin-orbit length}
\label{sec:spinorbitlength}
We now turn to the evaluation of the strength of the spin-orbit interaction from the measurements shown in Fig.~\ref{fig4}. The model developed in the previous sections facilitates the extraction of this strength from the width of the two leakage current features as a function of detuning in Fig.~\ref{fig4}(a)-(c). This width is given by the sizes $2\Delta_{\rm ST}^{\pm}$ of the avoided crossings (see Fig.~\ref{fig3}d) induced by the spin-orbit interaction. Here, the spin-flip tunneling energies $\Delta_{\rm ST}^{\pm}$ are functions of the spin-orbit length $\lambda_{so}$ and furthermore depend on the overlap of the wave functions of the $|T_{\uparrow\uparrow,\downarrow\downarrow}(1,1)\rangle$ states with those of the $|S_\pm\rangle$ states, as well as on the dot size. The spin-flip tunneling energy can be written as (see Appendix~\ref{app:C} for the complete derivation)
\begin{equation}
\label{eq:DeltaST}
\Delta_{\rm ST}^{\pm}=t_c \tan\left(\frac{a}{\lambda_{so}}\right) \sqrt{\frac{1\pm \cos(\theta)}{2}} \ ,
\end{equation}
with $\theta$ the mixing angle of the $|S_\pm\rangle$ states.\\
\begin{figure}[t]
	\includegraphics[width=0.50\textwidth]{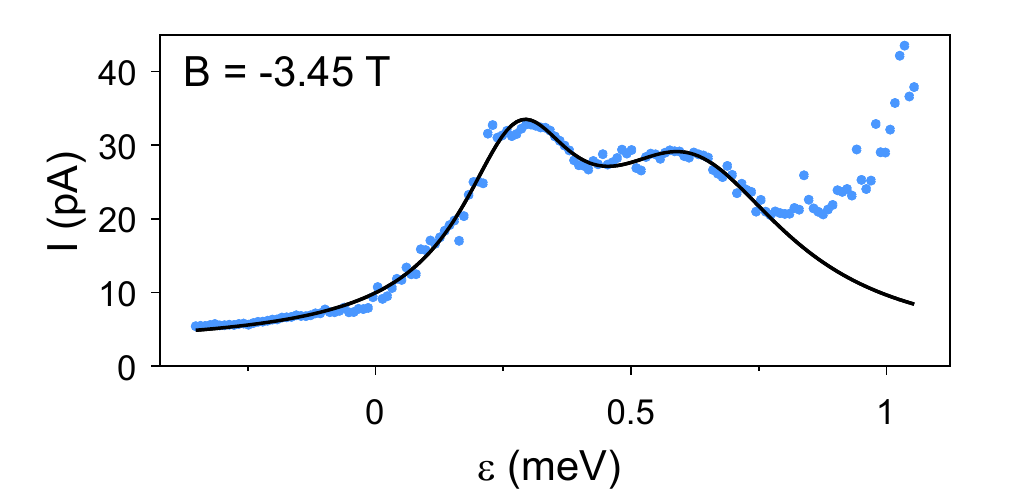}
	\caption{Measured leakage current as a function of detuning, for $V_\text{g3}$ = \SI{3820}{mV} and $B$ = \SI{-3.45}{T}. The black curve is a fit of Eq.~\eqref{eq:curr} to the data.}
	\label{fig5}
\end{figure}

The leakage current $I_{\pm}(B)$ corresponding to the resonances around $\varepsilon=\varepsilon_{\pm}(B)$ can be written as~\cite{vanderWielElectrontransportdouble2002,NazarovQuantuminterferencetunnel1993,StoofTimedependentresonanttunneling1996,LiPauliSpinBlockade2015}
\begin{equation}\label{eq:curr}
I_{\pm}=I_0+e \Gamma \frac{\left(\Delta_{\rm ST}^{\pm}\right)^2}{(\varepsilon-\varepsilon_{\pm})^2+3\left(\Delta_{\rm ST}^{\pm}\right)^2+h^2\Gamma^2/4} \ .
\end{equation}
Here, the lead-to-dot relaxation rate $\Gamma\sim 0.45 \text{ GHz}$ is taken to be symmetric for both of the leads and is estimated by adjusting the formula in Eq.~\eqref{eq:curr} for the $|S(0,2)\rangle\rightarrow |S(1,1)\rangle$ transition, and fitting it to the current measured for opposite $V_\text{SD}$. The offset current term $I_0$ contains all incoherent relaxation mechanisms, as well as $|S\rangle - |T_0\rangle$ mixing. Discussing this term in detail is beyond the scope of this paper and we refer the interested reader to Ref.~\cite{DanonPaulispinblockade2009}. Additionally, we note that since we operate at relatively high temperature, it might be expected that the transitions are thermally broadened. However, the temperature of \SI{1.4}{\kelvin} is still low compared to the orbital level splitting of \SI{1}{\milli\electronvolt}, making such broadening negligible. The dot-lead tunneling rate $\Gamma$ is influenced by temperature, but the value of $\Gamma$ that we determine independently from the measurements already includes this effect.

We therefore conclude that the width of the two leakage current features is given by the spin-flip tunneling energies $\Delta_{\rm ST}^{\pm}$, which are then deduced by fitting the Lorentzians in Eq.~\eqref{eq:curr} to the data sets of Fig.~\ref{fig4}(a)-(c). An example of this is shown in Fig.~\ref{fig5}. The color plots of Fig.~\ref{fig4}(d)-(f) are constructed from the Lorentzians found in this way for different values of the magnetic field. It can be seen that the model plots accurately reproduce the leakage current observed in the corresponding experimental data.\\

Importantly, the determined $\Delta_{\rm ST}^{\pm}$ allow to extract the spin-orbit length $\lambda_{so}$. Using Eq.~\eqref{eq:DeltaST}, we obtain the ratio $\lambda_{so}/a$ directly from the ratio $\Delta_{\rm ST}/t_c$ of the average spin-flip tunneling $\Delta_{\rm ST}=\big[(\Delta_{\rm ST}^+)^2+(\Delta_{\rm ST}^-)^2\big] ^{1/2}$ and the spin-conserving tunneling $t_c$.
This yields ratios of $\lambda_{so}/a$ as shown in Table~\ref{tab} for the different configurations of our double quantum dot. The precise value of the interdot distance $a$ cannot be exactly determined from the measurements, but we can roughly estimate $a\sim \SI{90}{\nm}$ by considering the distance between the gates $g_2$ and $g_4$ (see Fig.~\ref{fig1}(a)). Using this value, we obtain an average estimated value $\lambda_{so}\sim \SI{65}{\nm}$ for the spin-orbit length, with small variation between the measurements of Fig.~\ref{fig4}(a)-(c).\\

Together with the orbital effects of the magnetic field, this notably small $\lambda_{so}$ leads to a dependence of the $g$-factor on the magnetic field, as described by Eq.~\eqref{eq:g-factor}. This effect is large, since the spin-orbit length $\lambda_{so}$ and the confinement length along the wire $l_\parallel$ are of the same order of magnitude. In our measurements, this manifests itself in the additional bending of the transitions $\varepsilon_{\pm}(B)$ at high values of the magnetic field.

\begin{table}
	\vspace{1em}
	\centering
	\begin{tabular}{| >{\centering}m{1.3cm} | >{\centering}m{1.0cm} | >{\centering}m{0.8cm} | >{\centering}m{0.8cm} | >{\centering}m{1.0cm} | >{\centering}m{0.8cm} | >{\centering}m{0.8cm} | c |}
		\hline 
		&\T $V_\text{g3}$ & $\tilde{B}$ & $B_0$ & $t_c(\tilde{B})$ & $g(\tilde{B})$ & $\l_\parallel$ & $\lambda_{so}/a$ \B\\
		& (mV) & (T) & (T) & ($\mu$eV) &  & (nm) & \B\\
		\hline 
		\hline 
		\T Fig.~\ref{fig4}(a) & 3820 & 1.2 & 3.8 & 44 & 0.9 & 45 & 0.78\B\\
		\T Fig.~\ref{fig4}(b) & 3830 & 0.8 & 4.8 & 33 & 1.0 & 41 & 0.72\B\\
		\T Fig.~\ref{fig4}(c) & 3840 & 0.35 & 5.0 & 16 & 1.1 & 39 & 0.71\B\\
		\hline 
	\end{tabular} 
	\caption{Extracted hole spin parameters, obtained for the three datasets shown in Fig.~\ref{fig4} by fitting the model to the data as described in the main text.}
	\label{tab}
\end{table} 

\section{Conclusions and outlook}
Summarizing, we have characterized the strength of spin-orbit interaction for hole spins confined in a double quantum dot in a Ge/Si nanowire, using spectroscopy measurements in Pauli spin blockade. We found the spin-orbit length to be of the same order of magnitude as the dot length and interdot distance. This has the remarkable consequence that the $g$-factor exhibits a non-linear dependence on magnetic field, which we observe experimentally at high values of the magnetic field. 

The observation of this strong spin-orbit interaction in Ge/Si nanowires forms the starting point of various subsequent experiments in this material system. From the value of $\lambda_{so}$ we can estimate the Rabi frequency for electric dipole induced spin resonance~\cite{GolovachElectricdipoleinducedspinresonance2006, BulaevElectricDipoleSpin2007} mediated through spin-orbit interaction to be in the range of $\sim0.1-1$\,GHz, for realistic values of microwave amplitudes. Such Rabi frequencies form an excellent basis for the implementation of fast hole spin qubits in this system. 

Further characterization studies of the spin-orbit interaction in this platform are of interest, in particular because here a quantitative comparison to relevant theoretical works~\cite{KloeffelStrongspinorbitinteraction2011, MaierTunablefactorphononmediated2013, KloeffelCircuitQEDholespin2013, MaierMajoranafermionsGe2014, KloeffelDirectRashbaspinorbit2018} is challenging, due to the relatively high dot occupation number. For instance, direct Rashba spin-orbit interaction is predicted to lead to a profound dependence of the spin-orbit interaction as well as the $g$-factor on electric fields. While we observe a dependence of the $g$-factor on a gate voltage (see Table~\ref{tab}), a more complete investigation of these effects would include measurements of the strength of the spin-orbit interaction as function of electric field amplitude or orientation of magnetic field. Such tunability of $g$-factor and spin-orbit strength could enable individual addressability of spin qubits in coupling them to microwave fields, as well as provide a way to limit the impact of charge noise on spin coherence.

\section*{Acknowledgments}
We thank C. Kloeffel for helpful discussions. We acknowledge the support of the Swiss National Science Foundation (Ambizione Grant Nr. PZOOP2161284/1 and Project Grant Nr. 179024), the Swiss Nanoscience Institute (SNI), the EU H2020 Microkelvin Platform EMP (Grant Nr. 824109), the NCCR Quantum Science and Technology (QSIT), the Georg H. Endress Foundation, the EU FET (TOPSQUAD, Grant Nr. 862046), the EU H2020 research and innovation programme (Grant Nr. 862046), and the Netherlands Organization for Scientific Research (NWO).

\appendix
\section{Model Hamiltonian}
\label{app:A}
\noindent Here, we provide a more detailed analysis of the theoretical model used in the main text. 
The relevant physics of a single hole confined in a quantum dot can be captured by the effective 2-dimensional Hamiltonian
\begin{equation}
\label{eq:single-dot-H}
H=H_o+H_{so} + H_Z \ ,
\end{equation}
with 
\begin{subequations}
	\begin{align}
	H_o&= \frac{\pi_x^2}{2m_\perp}+\frac{\pi_y^2}{2m_\perp}+ \frac{\pi_z^2}{2m_\parallel} +\frac{m_\parallel\omega_\parallel^2}{2}z^2\nonumber\\
	&\;\;\;\;+\frac{m_\perp\omega_\perp^2}{2}(x^2+y^2) \ , \\
	H_{so} &= \alpha \pi_z \sigma_y \ ,\\
	H_Z &=\frac{g_0\mu_B}{2} \textbf{B}\cdot\pmb{\sigma} \ .
	\end{align}
\end{subequations}
Here, we define the dynamical momentum $\pmb{\pi}=-i\hbar \nabla-e\textbf{A}$, where $\textbf{A}$ is the vector potential accounting for an externally applied magnetic field $\textbf{B}$. These operators satisfy the commutation relations $[\pi_i,\pi_j]=i \epsilon_{ijk}\hbar e B_k$, $[\pi_i,x_j]=-i \hbar \delta_{ij}$. We model the confinement potential by an anisotropic harmonic oscillator, with confinement frequencies $\omega_\perp$ and $\omega_\parallel$, and effective masses $m_\perp$ and $m_\parallel$ in the direction perpendicular and parallel to the nanowire growth direction, respectively. In the following, we assume $\omega_\perp>\omega_\parallel$.
Because of the magnetic field, the spin states are split in energy by the Zeeman energy; here $g_0$ is the $g$-factor of the system and the field $\textbf{B}$ is assumed to be homogeneous. 
The interaction between different spin states is captured by a Rashba-like spin-orbit interaction $H_{so}$~\cite{KloeffelDirectRashbaspinorbit2018,KloeffelStrongspinorbitinteraction2011}.\\

\noindent Our final goal is to extract from the measurements the spin-orbit interaction parameter $\alpha$. It is convenient to introduce the spin-orbit length
\begin{equation}
\lambda_{so}= \frac{\hbar}{m_\parallel\alpha} \ ,
\end{equation}
and to perform the unitary spin-dependent displacement of states~\cite{LevitovDynamicalspinelectriccoupling2003}
\begin{equation}
\label{eq:Rashba-Levitov-unitary}
S=e^{i \sigma_ y z/\lambda_{so} }\ ,
\end{equation}
that diagonalizes the spin-orbit interaction in spin-space
\begin{equation}
S\Big(H_o+H_{so}\Big)S^{\dagger}=H_o-\frac{\hbar^2}{2m_\parallel \lambda_{so}^2} \ ,
\end{equation}
converting  the Zeeman term to a position-dependent quantity. We now focus on the case where the magnetic field points in the $x$-direction, i.e. $\textbf{B}=B \textbf{e}_x$, and we obtain
\begin{equation}
\label{eq:Zeeman-rotated}
S H_ZS^\dagger= \frac{g_0 \mu_B}{2} B \left[ \sigma_x \cos\!\left(\frac{2z}{\lambda_{so}}\right)+\sigma_z \sin\!\left(\frac{2z}{\lambda_{so}}\right) \right] \ .
\end{equation}\\

\noindent In the harmonic confinement approximation, the orbital Hamiltonian $H_o$ can always be diagonalized exactly. Assuming $B>0$, we can introduce the vector of gauge-independent canonical positions $\textbf{Q}$ and momenta $\textbf{P}$
\begin{equation}
\label{eq:original-coordinates}
\textbf{Q}=\left(\begin{array}{c}
\frac{z}{l_B}-\frac{l_B}{\hbar}\pi_y \\
\frac{l_B}{\hbar}\pi_y\\
x
\end{array}
\right) \ \ \ \text{and} \ \ \ \textbf{P}=\left(\begin{array}{c}
\frac{y}{l_B}+\frac{l_B}{\hbar}\pi_z \\
\frac{l_B}{\hbar}\pi_z\\
-i\partial_x
\end{array}
\right) \ ,
\end{equation}
satisfying $[Q_i,P_j]=i\delta_{ij}$; here $l_B=\sqrt{\hbar/(e|B|)}$ is the magnetic length. When $B<0$, the first two positions and momenta are swapped.
The coupled harmonic oscillators can be decoupled by the symplectic Bogoliubov  transformation 
\begin{equation}
\label{eq:symplectic-transf}
\left(\begin{array}{c}
\textbf{Q}\\
\textbf{P}
\end{array}
\right) =\left(\begin{array}{cc}
\mathcal{A}(r) & 0 \\
0 & \mathcal{A}(-r)^T
\end{array}
\right)\left(\begin{array}{c}
\textbf{q}\\
\textbf{p}
\end{array}\right)\ ,
\end{equation}
where 3-dimensional matrix $\mathcal{A}(r)$ is defined by
\begin{equation}
\mathcal{A}(r)=\left(
\begin{array}{ccc}
\cosh(r) & -\frac{\omega_\perp}{\omega_\parallel} \sinh(r) &0 \\
-\frac{\omega_\parallel}{\omega_\perp}\sinh(r) & \cosh(r)&0 \\
0&0&1
\end{array}
\right) \ ,
\end{equation}
with squeezing parameter
\begin{equation}
r=\frac{1}{2}\text{arccoth}\left(\frac{\frac{e^2 B^2}{m_\perp m_\parallel}+\omega_\perp^2+\omega_\parallel^2}{2\omega_\perp\omega_\parallel}\right) \ .
\end{equation}
In the new coordinate system with positions $\textbf{q}$ and momenta $\textbf{p}$, we obtain three independent harmonic oscillators with frequencies $\omega_\perp$ and $\omega_1<\omega_2$, where the Fock-Darwin frequencies are 
\begin{align}
\label{eq:symplectic-freq}
\omega_1&=\omega_2\tanh(r)\nonumber\\
&= \frac{\sqrt{m_\parallel m_\perp}}{eB}\omega_\parallel\omega_\perp\sqrt{\left(1-\frac{\omega_\parallel}{\omega_\perp}\tanh(r)\right)\left(1-\frac{\omega_\perp}{\omega_\parallel}\tanh(r)\right)} \ .
\end{align}
We point out that when $B\rightarrow 0$, Eq.~\eqref{eq:symplectic-freq} is still valid and it leads to the expected result $\omega_1=\omega_\parallel$ and $\omega_2=\omega_\perp$.\\

\noindent The groundstate $|0\rangle$ is the state simultaneously annihilated by the annihilation operators in this coordinate system
\begin{equation}
\label{eq:annhilation-op}
a_j=\frac{1}{\sqrt{2}}\left(\beta_j q_j+\frac{i}{\beta_j}p_j\right) \ ,
\end{equation}
where 
\begin{equation}
\beta_j=\left[\left(\frac{\omega_\parallel m_\parallel}{\omega_\perp m_\perp}\gamma\right)^{1/4} , \ \left(\frac{\omega_\parallel m_\parallel}{\omega_\perp m_\perp}\frac{1}{\gamma}\right)^{1/4} , \ \sqrt{m_\perp \omega_\perp/\hbar}\ \right]_j \ ,
\end{equation}
and $\gamma=\frac{\omega_\parallel}{\omega_\perp}\frac{\omega_\parallel/\omega_\perp-\coth(r)}{\omega_\perp/\omega_\parallel-\coth(r)}$.
To determine the groundstate wavefunction in real-space, we need to specify a gauge. In the symmetric gauge $\textbf{A}=\textbf{B}\times \textbf{r}/2$, and combining Eqs.~\eqref{eq:original-coordinates}, \eqref{eq:symplectic-transf} and  \eqref{eq:annhilation-op}, we obtain
\begin{equation}
\label{eq:GS-wavefunction}
\psi_{0}(\textbf{r})=\frac{1}{\pi^{3/4}\sqrt{l_x l_y l_z}}e^{-\frac{1}{2}\left(\frac{x^2}{l_x^2}+\frac{y^2}{l_y^2}+\frac{z^2}{l_z^2}\right)+i \frac{yz}{2l_B^2}\left(\frac{\omega_\parallel-\omega_\perp}{\omega_\parallel+\omega_\perp}\right)} \ ,
\end{equation}
where we defined the magnetic field-dependent lengths
\begin{equation}
\label{eq:lengths-B-field}
l_y=l_\perp \left(1+\frac{B^2}{B_0^2}\right)^{-1/4} \ \ \ \text{and} \ \ \ l_z=l_\parallel\left(1+\frac{B^2}{B_0^2}\right)^{-1/4} \ ,
\end{equation}
and the usual harmonic lengths
\begin{equation}
l_x=l_\perp =\sqrt{\frac{\hbar}{m_\perp \omega_\perp}} \ \ \ \text{and} \ \ \ l_\parallel=\sqrt{\frac{\hbar}{m_\parallel \omega_\parallel}}\ .
\end{equation}
The characteristic magnetic field $B_0$ in Eq.~\eqref{eq:lengths-B-field} determines the relevant field at which the orbital effects start to become significant and it is defined by
\begin{equation}
B_0=\frac{\sqrt{m_\parallel m_\perp}}{e}(\omega_\parallel+\omega_\perp) \ .
\end{equation}\\

\noindent Projecting the Hamiltonian in Eq.~\eqref{eq:single-dot-H} onto the groundstate subspace and subtracting a constant energy term, we obtain the effective low energy Hamiltonian
\begin{equation}
\label{eq:g-fator_GS}
H_{GS}=  \frac{g\mu_B B}{2}  \sigma_x \ ,
\end{equation}
where we introduce the effective $g$-factor
\begin{equation}
\label{eq:g-factor-renorm}
g=g_0e^{-l_z^2/\lambda_{so}^2} \ .
\end{equation}
We emphasize that the $g$-factor is renormalized by the spin-orbit interaction, and it acquires a magnetic field dependence via $l_z$, see Eq.~\eqref{eq:lengths-B-field}. \\

\noindent We remark that because of the transformation in Eq.~\eqref{eq:Rashba-Levitov-unitary}, we are now treating spin-orbit interaction exactly, and the perturbation coupling different orbital states comes from the space-dependent magnetic field in the Zeeman energy, see Eq.~\eqref{eq:Zeeman-rotated}. This approach is the most convenient to describe the results of this experiment, where a strong spin-orbit interaction is measured. Because of this term, the orbital ground state is coupled to the first excited orbital state $|1\rangle$  with energy $\hbar\omega_1$. 
In particular, the interaction is 
\begin{equation}
\langle 0|H|1\rangle=\frac{ l_B}{\sqrt{2}\lambda_{so}\beta_1}\left( \cosh(r)-\frac{\omega_\parallel}{\omega_\perp}\sinh(r)\right)g \mu_B B \sigma_z \ .
\end{equation}
Using the values extracted in the main text, see Table~\ref{tab}, we find that the amplitude of this interaction term is $\sim 20\%$ of the energy gap $\sim \hbar\omega_1$ at the maximal field measured $B=8$ T. Consequently, in the following we focus on the ground state subspace only.

\section{Double-dot Hamiltonian}
\label{app:B}
\noindent We now construct the double-dot effective Hamiltonian by using the Hund-Mulliken method. To do so, we create an orthonormal basis of harmonic eigenfunctions whose center of mass is at the positions $z=\pm a/2$. Here, $a$ is the interdot distance. Following the conventional procedure, we find the overlap matrix between the orbital ground states of the two dots: $\mathcal{P}_{ij}=\langle \Psi_i|\Psi_j\rangle$, where 
\begin{align}
|\Psi\rangle = \Big(&T_z\big(-a/2\big)S^\dagger |\psi_{0}\uparrow\rangle , \ T_z\big(-a/2\big)S^\dagger |\psi_{0}\downarrow\rangle ,\nonumber\\
&T_z\big(+a/2\big)S^\dagger |\psi_{0}\uparrow\rangle ,\ T_z\big(+a/2\big)S^\dagger |\psi_{0}\downarrow\rangle\Big)
\end{align}
The magnetic translation operators are defined as $T_z(X)=e^{i X (\pi_z/\hbar+y/l_B^2)}$ and $\psi_0$ is the ground state wavefunction in Eq.~\eqref{eq:GS-wavefunction}. 
Importantly, because the unitary $S^\dagger$ in Eq.~\eqref{eq:Rashba-Levitov-unitary} is spin-dependent, here  $\mathcal{P}$ is a $4\times 4$ matrix. Explicitly, we find 
\begin{equation}
\mathcal{P}=\tau_0\sigma_0+ s \cos\left(\frac{a}{\lambda_{so}}\right)  \tau_x \sigma_0 + s \sin\left(\frac{a}{\lambda_{so}}\right)  \tau_y \sigma_y \ ,
\end{equation}
where $\tau_i$ are Pauli matrices acting on the different dots, $\sigma_i$ are acting on spins and we define the small parameter
\begin{equation}
\label{eq:s-overlap}
s=e^{-\frac{a^2}{4l_z^2}\left(1+\frac{(\omega_\perp-\omega_\parallel)^2}{4\omega_\perp\omega_\parallel}\frac{B^2}{B^2+B_0^2}\right)} \ . 
\end{equation}
Orthogonal and symmetric states $|O\rangle$ are constructed from the non-orthogonal states $|NO\rangle$ by the linear map $|O\rangle=|NO\rangle \mathcal{P}^{-1/2}$ and single-particle operators $H$ transform as $H_{O}=\mathcal{P}^{-1/2}H_{NO}\mathcal{P}^{-1/2}$. The generalization to two-body operators is straightforward. \\

\noindent For rather general double-dot confinement potentials, we find that the orbital Hamiltonian in the orthonormal basis has the form
\begin{equation}
H_{o}=  t_c   \tau_x \sigma_0 + t_{so}  \tau_y \sigma_y +\frac{\varepsilon}{2}\tau_z\sigma_0 \ .
\end{equation}
Here, $\varepsilon$ is the detuning between the two dots typically caused by an electric field along the wire, $t_c$ is the spin-conserving  tunneling energy and $t_{so}$ is the spin-flip tunneling energy caused by the spin-orbit interaction. In particular, we find that 
\begin{subequations}
\begin{align}
\label{eq:tunneling}
t_c&=\frac{s}{1-s^2}t_0\cos\left(\frac{a}{\lambda_{so}}\right)\ \ \ \text{and} \\
t_{so}&= \frac{s}{1-s^2}t_0\sin\left(\frac{a}{\lambda_{so}}\right) = t_c \tan\left(\frac{a}{\lambda_{so}}\right)\ .
\end{align}
\end{subequations}
where $t_0$ is a characteristic energy dependent on the details of the confinement potential and the leading magnetic field dependence of the tunneling energy is caused by the exponential dependence of the overlap $s$ on $B$, see Eq.~\eqref{eq:s-overlap}. \\

\noindent  Also, the Zeeman energy in the orthogonal basis is
\begin{equation}
H_Z=\frac{g \mu_B B}{2}\Big( g_1 \tau_0\sigma_x+ g_2 \tau_x \sigma_x+g_3 \tau_z\sigma_z \Big) \ ,
\end{equation}
where we introduce the dimensionless prefactors 
\begin{align}
g_1&= \frac{1+\sqrt{1-s^2}-2 s^2 \cos \left(\frac{a}{\lambda_{so} }\right)}{2-2 s^2}\nonumber\\
&\;\;\;\;+\frac{\left(1-\sqrt{1-s^2}\right) \cos \left(\frac{2a}{\lambda_{so} }\right)}{2-2 s^2}\nonumber\\
&=1 +\mathcal{O}(s^2) \ , \\
g_2&= \frac{1-\cos\left(\frac{a}{\lambda_{so} }\right)}{1-s^2} s\ ,  \\
g_3&=\frac{ s^2-\left(1-\sqrt{1-s^2}\right) \cos \left(\frac{a}{\lambda_{so} }\right)}{1-s^2} \sin \left(\frac{a}{\lambda_{so} }\right) =\mathcal{O}(s^2)\ .
\end{align}
Neglecting corrections of order $s^2$, we can discard the term proportional to  $g_3$, that couple the triplet states $T_{\uparrow\downarrow}(1,1)$ to the singlet state $S(1,1)$. The term proportional to $g_2$ arise when the spin-orbit interaction is large and cause interactions between the triplet $T_0(1,1)$ and the doubly-occupied singlet states $S(2,0)$ and $S(0,2)$. This term causes an extra resonant peak of the leakage current, however, in the present experiment the energy of this interaction is of a few microelectronvolts, much smaller than the contribution due to the spin-flip tunneling. Consequently, in the following, we will ignore it and consider only $H_Z\approx {g \mu_B B}\tau_0\sigma_x/2$.\\

\noindent Coulomb interactions are also required to understand the physics of the system. In particular, the most relevant electrostatic interaction element for the current experiment is the addition energy, 
\begin{align}
\label{eq:U-charging}
U&=\langle\Psi_i\Psi_i|\frac{e^2}{4\pi \epsilon_s r}|\Psi_i\Psi_i\rangle\nonumber\\
&=\frac{e^2}{4\pi \epsilon_s}\sqrt{\frac{2}{\pi }}\frac{ F\left(\cos^{-1}\left(\frac{l_z}{l_x}\right)|\frac{l_x^2-l_y^2}{l_x^2-l_z^2}\right)}{\sqrt{l_x^2-l_z^2}} \ ,
\end{align}
where $F(a|b)$ is the elliptic $F$ function and $\epsilon_s=16\epsilon_0$ is the dielectric constant of germanium times the vacuum permittivity $\epsilon_0$. Eq.~\eqref{eq:U-charging} holds for general values of lengths $l_i$ provided that the appropriate limit is taken carefully. The next largest Coulomb interaction elements are the Hartree and Fock terms $U_{H}=\langle\Psi_i\Psi_{j\neq i}|\frac{e^2}{4\pi \epsilon_s r}|\Psi_{j\neq i}\Psi_i\rangle$ and $U_F=\langle\Psi_i\Psi_{j\neq i}|\frac{e^2}{4\pi \epsilon_s r}|\Psi_i\Psi_{j\neq i}\rangle$, respectively. In the present experiment, the overlap $s$ between wave functions of different dots is expected to be small, and so we discard the corrections of order $\mathcal{O}(s^2)$ and we ignore the exchange interaction $U_F\approx 0$. 

\section{Singlet-Triplet basis}
\label{app:C}
\noindent We can now rewrite the Hamiltonian in the singlet-triplet basis. Neglecting higher orbital states, the relevant triplet states are
\begin{align}
&||T_{\uparrow\uparrow,\downarrow\downarrow}(1,1)\rangle=  c_{-,\uparrow(\downarrow)}^\dagger c_{+,\uparrow(\downarrow)}^\dagger|0\rangle   \ \ \ \text{and}\nonumber\\ \ \ \  &|T_0(1,1) \rangle=  \frac{c_{-,\uparrow}^\dagger c_{+,\downarrow}^\dagger+c_{-,\downarrow}^\dagger c_{+,\uparrow}^\dagger}{\sqrt{2}} |0\rangle  \ ,
\end{align}
and the singlets are
\begin{align}
&|S(0,2) \rangle=  c_{+,\uparrow}^\dagger c_{+,\downarrow}^\dagger|0\rangle     \ \ \ \text{and}\nonumber\\ \ \ \   &|S_0(1,1) \rangle= \frac{c_{-,\uparrow}^\dagger c_{+,\downarrow}^\dagger-c_{-,\downarrow}^\dagger c_{+,\uparrow}^\dagger}{\sqrt{2}} |0\rangle  \ ,
\end{align}
where we introduce the fermionic ladder operators $c_{i,\sigma}^\dagger$ creating an electron at the $i$th dot with spin $\sigma$. We do not consider here the singlet state $S(2,0)$ because it is far detuned in energy, and so the interactions of these states with it are suppressed by the large energy difference.\\

\noindent By aligning the spin quantization axis to the direction of the magnetic field, we find in the singlet-triplet basis $\big(S(0,2),S(1,1),T_{\uparrow\uparrow}(1,1),T_{\downarrow\downarrow}(1,1),T_0(1,1)\big)^T$
\begin{equation}
H=\left(
\begin{array}{ccccc}
U-\varepsilon & \sqrt{2} t_c & t_{so} & -t_{so} & 0 \\
\sqrt{2} t_c  & U_H & 0 & 0 & 0 \\
t_{so} & 0 & U_H+ g \mu_B B &0 &0 \\ 
-t_{so} & 0 & 0& U_H-g \mu_B B  &0 \\ 
0 & 0 & 0 &0 & U_H
\end{array}
\right) \ ,
\end{equation} 
where $t_c$, $t_{so}$ and $g$ and $U$ are defined in Eqs.~\eqref{eq:tunneling},   \eqref{eq:g-factor-renorm} and \eqref{eq:U-charging}, respectively.
The singlet sector is hybridized by the spin-conserving tunneling energy. By introducing the hybridized singlet states $S_\pm$ obtained by rotating the singlet sector by $\theta/2$, where $\theta$ is
\begin{equation}
\theta=\arctan\left(\frac{2\sqrt{2} t_c}{U-U_{H}-\varepsilon}\right) \ ,
\label{Eq:theta}
\end{equation}
we can rewrite the Hamiltonian in the convenient form
\begin{equation}
H=\left(
\begin{array}{ccccc}
E_+ & 0 & \Delta_{\text{ST}}^+ & -\Delta_{\text{ST}}^+ &0 \\
0 & E_- & -\Delta_{\text{ST}}^- & \Delta_{\text{ST}}^- & 0 \\
\Delta_{\text{ST}}^+ & -\Delta_{\text{ST}}^- & U_{H}+g\mu_B B & 0 & 0 \\
-\Delta_{\text{ST}}^+ & \Delta_{\text{ST}}^- &0& U_{H}-g\mu_B B  & 0 \\
0& 0& 0& 0& U_H
\end{array}
\right) \ ,
\end{equation} 
where we defined the hybridized singlet energies $E_\pm$ and  the spin-orbit interaction $\Delta_{\text{ST}}^\pm$ via
\begin{subequations}
	\begin{align}
	\label{eq:energy_pm}
	E_\pm &=\frac{1}{2} \left(U+U_{H}-\varepsilon \right) \pm \sqrt{2 t_c^2+\frac{1}{4} \left(U-U_H-\varepsilon\right)^2} \ , \\
	\Delta_{\text{ST}}^\pm &=t_{so}\sqrt{\frac{1\pm \cos(\theta)}{2}} \ .
	\end{align}
\end{subequations}\\
Note that in the limit of weak spin orbit coupling, i.e., $a/\lambda_{so} \ll 1$, we recover the result obtained previously for the ST splitting~\cite{StepanenkoSinglettripletsplittingdouble2012}.

The leakage current is related to the matrix elements $\Delta_{\text{ST}}^\pm$ between singlet and triplet states via~\cite{NazarovQuantuminterferencetunnel1993,LiPauliSpinBlockade2015,StoofTimedependentresonanttunneling1996,vanderWielElectrontransportdouble2002}
\begin{equation}
I_\pm=e \Gamma_L \frac{ \big(\Delta_{\text{ST}}^\pm\big)^2}{(\varepsilon-\varepsilon_\pm)^2+\big(\Delta_{\text{ST}}^\pm\big)^2 \left(\frac{\Gamma_L}{\Gamma_R}+2\right)+{h^2{\Gamma_L}^2}/{4}} \ .
\end{equation}
where $\Gamma_{R(L)}$ is the coupling between the right, occupied (left, unoccupied) dot to the metallic lead and $\varepsilon_\pm$ is the position of the triplet $T_{\uparrow\uparrow,\downarrow\downarrow}(1,1)$ and the singlet $S_\pm$ anticrossing. In particular, by using Eq.~\eqref{eq:energy_pm}, we find
\begin{equation}
\varepsilon_\pm=U-U_{H}\pm \left( \frac{2t_c^2}{g\mu_B B}-g\mu_B B\right) \ .
\end{equation}
Neglecting the corrections due to the Hartree energy $U_H$, small compared to the addition energy $U$, and assuming symmetric dot-lead coupling $\Gamma_L\approx\Gamma_R=\Gamma$, we obtain Eqs.~\eqref{eq:epsa} and \eqref{eq:curr} of the main text. Note that in the main text the detuning is measured from the singlet-singlet anti-crossing, therefore Eq.~\eqref{eq:epsa} contains a constant energy shift.

\bibliography{200605Referencesbetterbibtex}

\end{document}